\documentclass[12pt]{article}

\usepackage{amsmath}

\usepackage{amsfonts}

\usepackage{amssymb}

\usepackage[latin1]{inputenc}

\usepackage{graphics}

\usepackage{graphicx}

\usepackage{epsfig}

\usepackage{amsthm}

\usepackage{lineno}

\usepackage{mathrsfs}

\usepackage{latexsym}

\newcommand{\h}{\hspace{.5cm}}
\newenvironment{destaque}{\begin{quotation}\small\em}{\end{quotation}}

\begin{document}

\title{The Harmonic Oscillator in the Classical Limit of a Minimal-Length Scenario}
\author{{\bf T. S. Quintela Jr., J. C. Fabris\footnote{e-mail: julio.fabris@cosmo-ufes.org} \hspace{0.01mm} and J. A. Nogueira\footnote{e-mail: jose.nogueira@ufes.br}}\\
{\small \it Departamento de F\'{\i}sica}\\
{\small \it Universidade Federal do Esp\'{\i}rito Santo}\\
{\small \it 29.075-910 - Vit\'oria-ES - Brasil}}

\maketitle

\begin{abstract}
\begin{destaque}
\h In this work we explicitly solve the problem of the harmonic oscillator in the classical limit of a minimal-length scenario. We show that (i) the motion equation of the oscillator is not linear anymore because the presence of a minimal length introduces an anarmonic term and (ii) its motion is described by a Jacobi sine elliptic function. Therefore the motion is periodic with the same amplitude and with the new period depending on the  minimal length. This result (the change in the period of oscillation) is very important since it enables us to find in a quite simple way the most relevant effect of the presence of a minimal length and consequently traces of the Planck-scale physics. We show applications of our results in spectroscopy and gravity.\\
\\
{\scriptsize PACS numbers: 12.60.i, 03.65.Sq, 04.20.Cv, 03.65.Ca}\\
{\scriptsize Keywords: Minimal length; Harmonic oscillator; Deformed Poisson Bracket.}
\end{destaque}
\end{abstract}

\pagenumbering{arabic}

\date{}


\section{Introduction}
\label{introd}

Recently the study of theories in a minimal-length scenario has attracted great interest, most likely because almost all attempts to quantization of gravity seem to suggest the existence of a minimal length~\cite{Amati:1988tn,Veneziano:1989fc,Maggiore1,Maggiore2}. Furthermore, these minimal-length scenarios can be considered an effective description of the effects of quantum gravity~\cite{Hossenfelder1}. The idea of existence of a minimal length is not new. In 1930, Heisenberg already proposed its existence as a natural cut-off~\cite{krag1, krag2, Heisenberg}. However, as far we know, M. Bronstein was the first to realize that the quantization of gravity leads to existence of a minimal length~\cite{Bronstein}.

The existence of a minimal length is somehow connected with the violation of the Lorentz symmetry, and even to non-locality. The connection among these proposals depends strongly on specific models. There is a huge literature about the constraints on Lorentz violation and non-locality~\cite{1,2,3,4,5,jms,7,8}. All these effects are supposed to play a fundamental role at very high energy scales, even if they can imply the existence of traces in some macroscopical phenomena. Such low energy scale traces depend also on specific models. The observational/experimental search for such signatures of the existence of minimal length, as well as for Lorentz symmetry violation and non-locality, even if  very precise for our usual standard, imply until the moment only upper limit bounds.

It was only in 1994 that A. Kempf, G. Mangano and R. B. Mann initiated the development of the mathematical base of the quantum mechanics in a minimal-length scenario~\cite{Kempf:1994su}. Over the years, many works have been published about theories in a minimal-length scenario, the most of them in a quantum context~\cite{Mead2,Hossenfelder2,chang,Sprenger,nozari,Hossenfelder0,tawfik}. However, the study of a minimal-length scenario in a classical context is also very important, like in cosmology~\cite{tawfik,farag1,farag2,tawfik2,zeynali,vakili,battisti} and in the experimental search for signals of the existence of a minimal length~\cite{benczik1,quesne,scardigli} for instance. One of the reasons for the study in a classical context is that there is not a clear separation between the physics of short and long distance~\cite{Gross,Witten}. The generalized uncertainty principle (GUP)
\begin{equation}
\Delta x \geq \hbar \left(\frac{1}{\Delta p} + \beta \Delta p \right),
\end{equation}
which arises from the modified commutation relations that introduce a minimal length, exhibits the basic features of the UV/IR correspondence~\cite{chang1}. This UV/IR mixing in the GUP makes possible effects of short distance phenomena to be present at lower energies~\cite{chang2}.  Furthermore, if we are convinced that gravitational effects are considered when the Heisenberg's canonical uncertainty relations are replaced with generalized uncertainty relations which introduce a minimal length, gravitational effects will be also considered in a classical context when generalized commutation relations are replaced with deformed Poisson brackets~\cite{li}.

One of the most important system of the classical mechanics is the harmonic oscillator, not only because is one the most frequent problem found in the literature, but also because a lot of others systems can be reduced to a similar problem that of the harmonic oscillator. Hence in this work we propose to solve explicitly the problem of a simple harmonic oscillator in the classical limit of a minimal-length scenario obtained by replacing the generalized commutation relations with deformed Poisson brackets. We will show that the new equation of motion is not linear anymore and its new period depends on the minimal length. We will also show that the amplitude of motion does not change. The dependency of the period on a minimal length is an extremely important result because it may be used to provide an experimental upper bound for the minimal length and probe the Planck-scale physics, such as in spectroscopy methods for vibrational diatomic molecules and in periodic astronomical systems. Although possible solutions for the harmonic oscillator  in a minimal-length scenario can be found in literature~\cite{chang1,pedram1,pedram2,esguerra,bawaj}, differently from theirs our result clearly shows that the motion remains periodic in such a way that its physical properties can be easily recognized, mainly the physical effects arising from the introduction of a minimal length.

The rest of this paper is organized as follows. In section 2 we introduce the classical limit of a minimal-length scenario. In section \ref{HO} we explicitly solve the modified equation of motion and calculate the new period. We also study two special cases of initial conditions. In section \ref{applic} we show applications of our results and also provide a very rough estimate for the upper bound of the minimal-length value using two systems of very different scales of dimension: the vibrational frequency of a molecule of $\text{H}_{2}$ and the rate of decrease of the orbital period of a binary system of neutron stars. We present our conclusions in section \ref{Concl}.


\section{The Classical Limit}
\label{CL}

The introduction of a minimal length in a quantum mechanical approach can be accomplished through of modifying the canonical commutation relations. A variety of modification choices can be considered. We concern with the most usual of them, proposed by Kempf~\cite{Kempf:1994su,Kempf2:1997}, which is given by
\begin{equation}
	\label{rc1kempf}
	[\hat{x}_i,\hat{p}_j] := i\hbar \biggl[  \left(1 + \beta\hat{\textbf{p}}^2 \right) \delta_{ij} + \beta^{\prime}\hat{p}_{i}\hat{p}_{j} \biggr],
\end{equation}
\begin{equation}
	\label{rc2kempf}
	[\hat{p}_i,\hat{p}_j] := 0,
\end{equation}
\begin{equation}
	\label{rc3kempf}
	[\hat{x}_i,\hat{x}_j] = -i\hbar \biggl[ 2\beta - \beta^{\prime} + \left(2\beta + \beta^{\prime} \right)\beta\hat{\textbf{P}}^2 \biggr] \sum_{k=1}^{3} \epsilon_{ijk}\hat{L}_{k},
\end{equation}
where $\beta$ and $\beta^{\prime}$ are parameters related to the minimal length, $\hat{\textbf{p}}^{2} = \sum_{i=1}^{3} \hat{p}_{i}^{2}$ and
\begin{equation}
\label{ang_moment}
\hat{L}_{k} = \frac{\left(\hat{x}_{i}\hat{p}_{j} - \hat{x}_{j}\hat{p}_{i} \right)}{\left(1+ \beta \textbf{p}^2 \right)}
\end{equation}
are the components of the orbital angular momentum operator, satisfying the usual commutation relations\footnote{ Theories in a minimal length scenario lead to a non-linear relation between the linear momentum and the wave vector of a particle. It is possible to interpret this non-linear relation as a Planck's constant depending on the energy or momentum~\cite{Hossenfelder2,Hossenfelder0,chang1,calmet}.}. The Eq.~(\ref{rc3kempf}) is obtained from Eqs.~(\ref{rc1kempf}) and (\ref{rc2kempf}) by using the Jacobi identity. The commutation relation (\ref{rc1kempf}) implies the existence of a non-zero minimal uncertainty in the position $\Delta x_{min} = \hbar \sqrt{3 \beta + \beta^{\prime}}$.

It is worth noting that in general the components of the position operator do not commute anymore. It is also easy to see that the commutation relation between $\hat{x}_{i}$ and $\hat{x}_{j}$ is not invariant under space translations because of $\left(\hat{x}_{i}\hat{p}_{j} - \hat{x}_{j}\hat{p}_{i} \right)$. However, a quick glance at Eq.~(\ref{rc3kempf}) shows that the choice $\beta^{\prime} = 2 \beta$ recovers the space invariance and commutation in first-order of $\beta$.

The classical limit of a minimal-length scenario can be obtained by replacing commutators with Poisson brackets~\cite{chang,tawfik,benczik1,chang1},
\begin{equation}
    \label{dpb}
    \frac{1}{i \hbar} [\hat{A}, \hat{B}] \rightarrow \{A, B\}.
\end{equation}

The requirement that the Poisson bracket must retain its properties of antisymmetry, linearity, product law (Leibniz rule) and Jacobi identity leads to the general form
\begin{equation}
	\label{gpb}
	 \{A, B\} := \sum_{i, j} \left[ \left( \frac{\partial A}{\partial x_{i}} \frac{\partial B}{\partial p_{j}} - \frac{\partial A}{\partial p_{i}} \frac{\partial B}{\partial x_{j}} \right)
	\{ x_{i}, p_{j} \} + \frac{\partial A}{\partial x_{i}} \frac{\partial B}{\partial x_{j}}\{ x_{i}, x_{j} \} \right] ,
\end{equation}
for any two functions of the coordinates and momenta. Therefore, the time evolution of the coordinates and momenta is given by 
\begin{equation}
	\label{xte}
	\dot{x}_{i} = \{x_{i}, H\} = \sum_{j} \left[ \frac{\partial H}{\partial p_{j}} \{ x_{i}, p_{j} \} + \frac{\partial H}{\partial x_{j}}\{ x_{i}, x_{j} \} \right]
\end{equation}
and
\begin{equation}
	\label{pte}
	\dot{p}_{i} = \{p_{i}, H\} = - \sum_{j} \left[ \frac{\partial H}{\partial x_{j}} \{ x_{i}, p_{j} \} \right] ,
\end{equation}
where $H$ is the Hamiltonian of the system.

Note that we keep the parameters $\beta$ and $\beta^{\prime}$ fixed as $\hbar \rightarrow 0$, in the transition from quantum to classical approaches\footnote{ From a string theory point of view, that corresponds to keep the string momentum scale fixed while the string length scale is taken to zero~\cite{vakili,battisti,chang1}.} \cite{vakili,battisti,benczik1,tkachuk}.


\section{The Harmonic Oscillator}
\label{HO}

The Hamiltonian of a linear harmonic oscillator of $m$ mass and $\omega_{0}$ ``angular frequency'' is given by
\begin{equation}
	\label{hoh}
	H = \frac{p^2}{2m} + \frac{1}{2} m \omega_{0}^{2} x^{2}.
\end{equation}

In 1-dimensional case the Poisson bracket between position and momentum becomes\footnote{Note that we have absorbed $\beta^{\prime}$ into $\beta$. Also note that now $\Delta x_{min} = \hbar \sqrt{\beta}$.}
\begin{equation}
	\label{xppb}
	\{x, p\} = 1 + \beta p^{2}.
\end{equation}
Therefore the motion equations read as
\begin{equation}
	\label{xme}
	\dot{x} = \frac{p}{m}\left(1 + \beta p^{2} \right)
\end{equation}
and
\begin{equation}
	\label{pme}
	\dot{p} = -m \omega_{0}^{2}x\left(1 + \beta p^{2} \right).
\end{equation}

We can eliminate the $\left(1 + \beta p^{2} \right)$ factor by dividing Eq.~(\ref{xme}) by Eq.~(\ref{pme}),
\begin{equation}
	\label{xpdivpp}
	\frac{\dot{x}}{\dot{p}} = - \frac{p}{m^{2} \omega_{0}^{2} x},
\end{equation}
and through a reduction of order to get the equation
\begin{equation}
	\label{eq1}
	p^{2} = - m^{2} \omega_{0}^{2} x^{2} + A^{2},
\end{equation}
where $A$ is a constant\footnote{Of course, that was just as expected because Eq.~(\ref{xpdivpp}) can also be obtained deriving Eq.~(\ref{hoh}).}. We can find the $A$ constant by substituting Eq.~(\ref{eq1}) into the Hamiltonian (\ref{hoh}). Thus, $A = \sqrt{2mE}$, where $E$ is the energy of the system.

Now, performing the derivative of Eq.~(\ref{xme}) we have
\begin{equation}
	\label{x2p}
	\ddot{x} = \frac{\dot{p}}{m} \left(1 + 3\beta p^{2} \right).
\end{equation}
And substituting Eq.~(\ref{eq1}) into Eq.~(\ref{pme}) we get
\begin{equation}
	\label{pp}
	\dot{p} = - m \omega^{2}_{0} \left(1 + 2\beta m E \right)x + \beta m^{3} \omega^{4}_{0}x^{3}.
\end{equation}

In order to eliminate the momentum we substitute Eqs.~(\ref{eq1}) and (\ref{pp}) into Eq.~(\ref{x2p}). Hence, throwing out terms of ${\cal O}(\beta^{2})$, we obtain
\begin{equation}
	\label{mex2p}
	\ddot{x} = - \omega^{2}x + 2b^{2}x^{3},
\end{equation}
where\footnote{Note that $\ddot{x} + \omega^{2}_{0}x + \beta \left( 8mE\omega^{2}_{0}x - 4 \beta m^{2} \omega_{0}^{4} x^{3} \right) = 0$.} $\omega^{2} := \omega^{2}_{0} \left(1 + 8 \beta m E \right)$ and $b^{2} := 2 \beta m^{2} \omega_{0}^{4}$. We can see that the existence of a minimal length introduces an anarmonic term (and a change to the oscillator frequency).

Once again, through a reduction of order we obtain from Eq.~(\ref{mex2p})
\begin{equation}
	\label{x2}
	\dot{x}^{2} + \omega^{2}x^{2} - b^{2}x^{4} = C^{2},
\end{equation}
where $C$ is a constant, which can be determinate using Eqs.~(\ref{hoh}), (\ref{xme}) and (\ref{x2}) as  $C = \sqrt{\frac{2E}{m}} \left( 1 + 2 \beta m E \right)$.

After some algebra (factoring the expression under the square root), Eq.~(\ref{x2}) can be written into form
\begin{equation}
	\label{integral01}
	\int_{t_{1}}^{t} dt = \pm \frac{1}{b} \int_{x_{1}}^{x} \frac{dx}{\sqrt{\left( x^{2} - x^{2}_{+} \right) \left( x^{2} - x^{2}_{-} \right)}},
\end{equation}
where
\begin{equation}
	\label{omegas}
	x^{2}_{\pm} := \frac{\omega^{2}}{2b^{2}} \left[ 1\pm \sqrt{1 -\frac{4b^{2}C^{2}}{\omega^{4}}} \right].
\end{equation}
The above equation imposes on $C$ the condition
\begin{equation}
	\label{Ccondition}
	C \leq \frac{1 + 8 \beta mE}{m \sqrt{8 \beta}}.
\end{equation}

With the change of variable $x := x_{-}\sin{\theta}$ and taking $t_{1} = t_{eq}$, such that $x(t_{eq}) = 0$, we have
\begin{equation}
	\label{integral03}
	w = b x_{+} \left( t - t_{eq} \right) = \int_{0}^{\theta(t)} \frac{d\theta}{\sqrt{1 - k^{2}\sin^{2}{\theta}}},
\end{equation}
where $k := \frac{x_{-}}{x_{+}} < 1$.

The above integral is the incomplete elliptic integral of the first kind whose the amplitude $\theta(t)$ is given by $sin\left[\theta(t)\right] = sn[w,k]$, and $sn[w, k]$ is the Jacobi sine elliptic function. Therefore
\begin{equation}
	\label{xt1}
	x(t) = x_{-} sn\left[b x_{+} (t - t_{eq}), k \right]
\end{equation}
and
\begin{equation}
	\label{vt}
	{\dot x}(t) = b x_{-} \sqrt{x_{+}^{2} - x^{2}} cn\left[b x_{+} (t - t_{eq}), k \right].
\end{equation}

From Eq.~(\ref{xt1}) it is easy to see that the amplitude of the motion is $x_{max} = x_{-} = \sqrt{\frac{2E}{m \omega_{0}^{2}}}$, thus
\begin{equation}
	\label{xt}
	x(t) = \sqrt{\frac{2E}{m \omega_{0}^{2}}} sn\left[b x_{+} (t - t_{eq}), k \right],
\end{equation}
and from Eq.~(\ref{vt}) we obtain $\dot{x}_{max} = bx_{-}x_{+} = C = \sqrt{\frac{2E}{m}} \left( 1 + 2 \beta m E \right)$.

Moreover the equation for $p(t)$ can easily be obtained from Eq.~(\ref{hoh}) as
\begin{equation}
	\label{pt}
	p(t) = \sqrt{2mE} cn\left[b x_{+} (t - t_{eq}), k \right].
\end{equation}

The zeros of $sn[u,k] = 0$ occur for $u = 2n {\sl K}(k)$, where $n = 0, 1, 2, \dots$ and
$$
	{\sl K}(k) = \int_{0}^{\frac{\pi}{2}} \frac{d \theta}{\sqrt{1 - k^{2} \sin^{2}{\theta}}},
$$
is the complete elliptic integral of first kind.

The time between two zeros is the half period. Consequently, the period of the (harmonic) oscillator in a minimal-length scenario is given by
\begin{equation}
	\label{periodo}
	{\sl T} = \frac{4}{b x_{+}} \int_{0}^{\frac{\pi}{2}} \frac{d \theta}{\sqrt{1 - k^{2} \sin^{2}{\theta}}} =  \frac{4}{b x_{+}}{\sl K}(k).
\end{equation}

To zero-order in $\beta$, the results (\ref{xt}) and (\ref{periodo}) take the forms
\begin{equation}
	\label{xt0}
	x(t) = \sqrt{\frac{2E}{m \omega_{0}^{2}}} \sin \bigl[\omega_{0} \left(t - t_{eq}\right) \bigr]
\end{equation}
and
\begin{equation}
	\label{periodo0}
	{\sl T} = \frac{2 \pi}{\omega_{0}}.
\end{equation}
Hence, to zero-order in $\beta$ we recover the results known for the ordinary harmonic oscillator\footnote{We use the expression ordinary harmonic oscillator in opposition to ''harmonic'' oscillator in the presence of a minimal length.}, as we expected.

Since the period is modified by presence of a minimal length, it is instructive to write it out to first-order in $\beta$. Then
\begin{equation}
	\label{periodo1}
	{\sl T} = \frac{2 \pi}{\omega_{0}} \left( 1 - \beta mE \right).
\end{equation}

Above equation shows that the period of the motion is shorter than ordinary period. Therefore the presence of a minimal length induces a faster motion of the oscillator. Thus, if we are convinced that the presence of a minimal length is a consequence of the gravity, the decrease in the period  and the consequent faster motion arises as an effect of gravity.

When it is only considered the best relevant effect of the presence of a minimal length, that is, the change in the frequency of the harmonic oscillator, we can write
\begin{equation}
\label{xfreq}
x(t) = \sqrt{\frac{2E}{m \omega_{0}^{2}}} \sin \bigl[\omega_{ML} \left(t - t_{eq}\right) \bigr],
\end{equation}
where
\begin{equation}
\label{wml}
\omega_{ML} = \omega_{0} \left( 1 + \beta m E \right).
\end{equation}

\subsection{Initial conditions. Special cases}

The instant $t_{eq}$ can be found from initial conditions. We are going to consider two special cases:

\begin{itemize}
\item[(i)] if $x(0) = 0$ and $\dot{x}(0) = \dot{x}_{0}$ then $t_{eq} = 0$ and $C = {\dot x}_{0}$.\\
Moreover, we have $\dot{x}_{0} = \sqrt{\frac{2E}{m}} \left( 1 + 2\beta mE \right)$.\\
To first-order in $\beta$ we obtain $E = \frac{m \dot{x}_{0}^{2}}{2} \left( 1 - \beta m^{2} \dot{x}_{0}^{2} \right)$.
\item[(ii)] if  $x(0) = x_{0}$ and ${\dot x}(0) = 0$ then $t_{eq} = \frac{1}{b x_{+}} {\sl K}(k) = \frac{\sl T}{4}$ and $C = \omega_{0}x_{0} \left( 1 + \beta m^{2}\omega_{0}^{2}x_{0}^{2} \right)$.\\
Moreover, we have $p_{0} = 0$ and $x_{0} = x_{max} = \sqrt{\frac{2E}{m\omega_{0}^{2}}}$.
\end{itemize}

Note that the condition (\ref{Ccondition}) is satisfied in both above cases, as must be.


\section{Applications in Microscopic and Macroscopic Systems}
\label{applic}

Since the frequency of a oscillator is a classical characteristic, the previous results can be used for probing the Planck-scale physics in oscillating astronomical systems, as well as even in oscillating quantum systems. Thus, for instance, the total radiated power by an oscillating electric dipole is given by
\begin{equation}
\label{rdele}
\left\langle P \right\rangle = \frac{\omega^{4}_{ML} p^{2}_{0}}{12 \pi \epsilon_{0} c^{3}} = \left\langle P \right\rangle_{0} + \frac{\beta \omega^{6}_{0} p^{4}_{0} m^{2}}{6 \pi \epsilon_{0} q^{2} c^{3}},
\end{equation}
where $p_{0}$ is the maximum value of the dipole moment and $m$ the mass of the $q$ oscillating charge.

Due to the growing increase in the accuracy of the frequency measurement~\cite{jms} the change in the period of oscillation can allow spectroscopy methods lead to an improvement in the experimental constraints on the upper bound for the minimal-length value.  Recently, D. Bouaziz computed an upper bound for the minimal length of $10^{-12}$ m by considering the vibrational ground state energy of the hydrogen molecule~\cite{Bouaziz}. We can make an estimation of the magnitude this upper bound, even though it is only very crude. The vibrational frequency of a molecule of $\text{H}_{2}$ is of order $10^{14}$ Hz ($4.4 \times 10^{3}$  $\text{cm} ^{-1}$) and the vibrational energy around of $10^{-1}$ eV. Considering that the accuracy of the experimental measures is of order $10^{-4}$  $\text{cm}^{-1}$~\cite{dickenson, Niu}, we obtain to the upper bound value of the minimal length $l_{min} < 10^{-15}$ m. Although our result is bigger than one obtained from the ground state energy of the hydrogen atom, $10^{-17}$ m, calculated through the Schroedinger equation~\cite{brau} or the Dirac equation~\cite{antonacci}, it can be improved if eventually more precise measurements are taken into account.

Now, let us consider a system consisting of two neutron stars of same mass $M$ orbiting around the common center of mass~\cite{ryder}. Because both stars have equal masses they describe circular motions of same radius centred at the mass center, whose projections on the Cartesian axes $X$ and $Y$ execute simple harmonic motion\footnote{Although the radius decreases, the chirp time is very very long (it is of the order of the Universe age) such a way that $R$ can be considered a constant for relatively short time of observation.}. Hence, we assume that the effective effect of the presence of a minimal length on the binary system is revealed as a change in the frequency of the simple harmonic motion, that is, through the observed orbital period. Therefore we expect that the observed frequency of the orbital motion is
\begin{equation}
\label{fosb}
\omega_{ML} = \sqrt{\frac{GM}{4R^{3}}}\left( 1 + \beta \frac{G M^{3}}{8 R} \right),
\end{equation}
where $R$ is the radius of the orbit at the considered instant of time and $\omega_{0} = \sqrt{\frac{GM}{4R^{3}}}$ is the orbital frequency calculated theoretically. The radiated power by system is given by
\begin{equation}
\label{potsb}
\frac{dE}{dt} = \frac{128 G}{5 c^{5}}  M^2 R^4 \omega^{6}_{0} + 0.3 \beta \frac{ G^{5} M^{8}}{ c^{5} R^{6}}.
\end{equation}

More interesting is the estimation of the decrease of the orbital period. In according to our assumption, due to the presence of a minimal length the period is given by
\begin{equation}
\label{posb}
{\sl T_{ML}} = {\sl T_{0}} \left( 1 - \beta \frac{G M^{3}}{8R} \right).
\end{equation}
Then, the rate of change of the period is
\begin{equation}
\label{rposb}
\frac{d{\sl T_{ML}}}{dt} =  \frac{d{\sl T_{0}}}{dt} \left( 1 - \beta \frac{G M^{3}}{24R} \right) .
\end{equation}

In accord with the literature the accuracy concerning the measurement of $\frac{d{\sl T_{ML}}}{dt}$ is about $6.0 \times 10^{-15}$~\cite{Baryshev,taylor1,taylor2}. Thus, for a neutron star of mass 1.4 times the mass of the Sun and $R = 4.0 \times 10^{8}$ m, we obtain to the upper bound of the minimal-length value $\Delta x_{min} \leq 1.5 \times 10^{-71}$ m. In a similar way to the result of $2.3 \times 10^{-68}$ m obtained by Benczik et al.~\cite{benczik1} using the observed precession of the perihelion of Mercury our estimative is many orders of magnitude below the Planck length.

A neutron star is in fact a system made up of many particles  and this is the possible reason for that so small result. In reference~\cite{quesne}, C. Quesne and V. M. Tkachuk have made a proposal which recovers $2.4 \times 10^{-17}$ m for the upper bound value of the minimal length in the case of the perihelion of Mercury.

In according to reference~\cite{quesne} we have
\begin{equation}
\hbar \sqrt{\tilde{\beta}} \leq 1.5 \times 10^{-71} {\text m},
\end{equation}
where $\tilde{\beta}$ is the effective parameter related to the minimal length of the neutron star which is given by
\begin{equation}
\label{emlp}
\tilde{\beta} = N_{nucl} \beta_{nucl} \left( \frac{m_{nucl}}{M} \right)^{3} + N_{elect} \beta_{elect} \left( \frac{m_{elect}}{M} \right)^{3},
\end{equation}
where $N_{nucl}$ and $N_{elect}$ are the numbers of nucleons and electrons which make up the neutron star and $\beta_{nucl}$ and $\beta_{elect}$ are the parameters related to nucleons and electrons.  The number of nucleons of the neutron star can be easily calculated as $N_{nucl} = \frac{M}{m_{nucl}} = 1.7 \times 10^{57}$. Since nucleons are made up of three quarks $\beta_{quark} = 3^{2} \beta_{nucl}$. Moreover, electrons and quarks are fundamental particles so it is assumed that $\beta_{quark} = \beta_{elect}$. Because $m_{elect} \ll m_{nucl}$ the second term of right hand side of Eq.~(\ref{emlp}) is negligible. Consequently we find that $ \Delta x_{min}^{quark} \leq 7.5 \times 10^{-14}$ m.
  

\section{Conclusion}
\label{Concl}

In this work we have shown that the motion of an ordinary harmonic oscillator in a minimal-length scenario introduced by the Kempf algebra (\ref{rc1kempf}) is given by Eq.~(\ref{xt}), that is, by a Jacobi sine elliptic function. This has allowed us to conclude easily that it turn into an anarmonic oscillator in which the amplitude is not changed and the period modified by the presence of the minimal length is given by Eq.~(\ref{periodo}).

We have also found $\dot{x}(t)$ as a Jacobi cosine elliptic function. However its $\dot{x}_{max}$ maximum value is modified by the presence of the minimal length. Such result is expected because in a minimal-length scenario $\dot{x}$ is not proportional to $p$ linear momentum (which is proportional to a Jacobi cosine elliptic function, Eq.~(\ref{pt})).

Since the minimal length induces a non-linear term proportional to $x^{3}$ into equation of motion, the reader may be questioning about the stability of the system. Eq.~(\ref{mex2p}) shows that there are one stable-equilibrium point at $x = 0$ and two unstable-equilibrium points at $x^{'} = \pm \sqrt{\frac{2E}{m \omega_{0}^{2}} \left( 1 + \frac{1}{8 \beta mE} \right)} = \pm x_{max}\sqrt{ 1 + \frac{1}{8 \beta mE} }$. Thus, the maximum points increase with the oscillator energy in a way that $x_{max}$ is always smaller than $\| x^{'} \|$, which assures the stability of the system for any energy. 

The result the period depends on the minimal length suggests its use to provide an experimental upper bound for the minimal length and probe the Planck-scale physics. Moreover the result (\ref{periodo1}) shows that the change in the period can be significant for great values of mass and energy. This suggests that oscillating astronomical systems may be more appropriate for probing the Planck-sacle physics. Thus, comparing our result with the precision of the measurement of the rate of change of the orbital period of a binary system of neutron stars we estimate the upper bound of the minimal-length value as $\Delta x_{min} \leq 10^{-71}$ m. The reliable result $ \Delta x_{min}^{quark} \leq 7.5 \times 10^{-14}$ m is obtained when it is taken into account that a neutron star is a system made up of many particles~\cite{quesne}. Since the oscillator period is a classical characteristic the result (\ref{periodo1}) or (\ref{wml}) can be also used for probing the Planck-scale physics even in measuring the quantum-oscillator period. Using experimental data of  spectroscopy methods, we have roughly estimated the upper bound for the minimal-length value of the order of $10^{-15}$ m, which can be improve taking into account the more precise measurements.

P. Pedram has found the same change in the frequency of a harmonic oscillator~\cite{pedram1}, but S. Ghosh and P. Roy have found a reduction in the frequency by considering coherent states of the harmonic oscillator in the presence of a minimal length~\cite{ghosh}.  K. Nozari has shown that in the presence of a scenario of minimal length there is no harmonic oscillation in quantum optics and complete coherency is impossible, consequently it is impossible to have a monochromatic ray \cite{nozari2}. M. Bawaj et al. have measured the changes of oscillating frequencies with the amplitudes for micro and macro-oscillators (with masses of $10^{-4}$, $10^{-7}$ and $10^{-11}$ kg)~\cite{bawaj}. They have obtained upper bound for the minimal length many orders of magnitude below the commonly found in the literature, perhaps because the oscillators in their experiments in fact are composed of many particles. Finally, we point out that our result for the period in order $\beta$ is in agreement with the result of Chang et al.~\cite{chang1}.


\section*{Acknowledgements}

\h We would like to thank FAPES, CAPES and CNPq (Brazil) for financial support.
\\




\end{document}